\documentclass[twocolumn,showpacs,nofootinbib,superscriptaddress]{revtex4-1}
\usepackage[utf8]{inputenc}
\usepackage[T1]{fontenc}
\usepackage[bookmarks,hidelinks]{hyperref}
\usepackage{bookmark}
\usepackage[british]{babel}
\usepackage{comment}

\usepackage{amsmath,amsthm,amsfonts,relsize,bm} 
\usepackage{graphicx,float} 
\usepackage{etoolbox}
\usepackage[shortlabels]{enumitem}
\usepackage{xcolor}
\usepackage{centernot}

\newcommand{\ed}{\mathrm{d}}

\newcommand{\Riem}[3][R]{{#1}^{#2}_{\hspace{0.5em}#3}}

\newcommand{\Lag}{\mathcal{L}} 
\newcommand{\tr}[1]{\mathrm{tr}\left\{#1\right\}} 
\newcommand{\sqrtg}{\sqrt{\vert g\vert}}
\newcommand{\bbar}[1]{\bar{\bar{#1}}}

\begin{document}

\title{Cosmological Solutions of the Non-minimally Coupled Weyl Connection Gravity Theories}

\author{Rodrigo Baptista}
\email{rodrigo.baptista@fc.up.pt}
\affiliation{Departamento de Física e Astronomia, Faculdade de Ciências da Universidade do Porto, Rua do Campo Alegre 687, 4169-007 , Porto, Portugal}

\author{Orfeu Bertolami}
\email{orfeu.bertolami@fc.up.pt}
\affiliation{Departamento de Física e Astronomia, Faculdade de Ciências da Universidade do Porto, Rua do Campo Alegre 687, 4169-007 , Porto, Portugal}
\affiliation{Centro de Física do Porto, Rua do Campo Alegre 687, 4169-007 , Porto, Portugal}

\date{\today}

\begin{abstract}
	\vspace*{0.2cm}
	We consider a  non-minimally coupled curvature-matter gravity model (NMC) with a Weyl connection, a theory referred to as non-minimally coupled Weyl connection gravity (NMCWCG).
	The Weyl connection is an affine connection that is not compatible with the metric, and involves a vector field.
	Assuming a vacuum expectation value for the vector field and a matter Lagrangian that only contains the contributions of the vacuum energy, we show that the model admits solutions in the space-form with a reference curvature that can be fine-tuned to be much smaller than the contribution of the matter fields.
	This shows that, at least in principle, the model admits a viable cosmological description.
	\vspace*{0.8cm}
\end{abstract}
\maketitle

\section{Introduction}

In 1918, Hermann Weyl introduced what we today call the Weyl connection \cite{Weyl}.
This connection was a first attempt to unify the only two known interactions of the time: gravity, described by the Theory of General Relativity; and electromagnetism described by Maxwell's theory.
Although not successful in its original purpose, the idea of an affine connection not compatible with the metric and, in particular, one whose covariant derivative of the metric depends on a vector field has been recently re-examined \cite{Gomes-Bertolami}\footnote{Even though in Ref.~\cite{Gomes-Bertolami} the proposed model was referred to as `Weyl gravity', we prefer to use here `Weyl connection gravity' instead, so to avoid confusion with models involving the square of the Weyl tensor in the action.}.

On the other hand, another way to highlight the role of the Weyl connection is to consider it in the context of gravity models with non-minimal coupling between curvature and matter Ref.~\cite{Bertolami-NMC}. These NMC models
extend well-known $f(R)$-theories by introducing an additional function, $f_2(R)$, that couples curvature to matter.
This coupling introduces new features to the gravity theory: the energy-momentum tensor is no longer covariantly conserved; there is a deviation from geodetic motion, and there are several implications for cosmology.
For an in-depth description of the NMC model and its implications see Refs.~\cite{Bertolami-NMC,March-2019,Bertolami-2013,Bertolami-2009,Bertolami-2010,Bertolami-2011.1,Bertolami-2011.2,Bertolami-2012,Bertolami-2015} and references therein.
In the present work we aim to introduce the Weyl connection into the NMC model, in a theory dubbed as non-minimally coupled Weyl connection gravity (NMCWCG).
This model was originally discussed in Ref.~\cite{Gomes-Bertolami} without endowing the Weyl vector field with a kinetic term.

The purpose of this paper is to study the aforementioned theory and to examine if space-time admits, in this model, constant sectional curvature, i.e. a space-\linebreak-form, whose associated curvature is small in comparison with the vacuum contribution of matter fields.
This is an important step to show that the model allows for setting up a sensible cosmology.

\section{The Weyl Connection}

A Weyl connection introduces, through the covariant derivative $D_\mu$, a vector field $A_\mu$ such that:
\begin{equation}\label{def:WeylG-coord}
	D_\lambda g_{\mu\nu}=A_\lambda\,g_{\mu\nu},
\end{equation}
where
\begin{equation}\label{WeylConnection-LCConnection}
	D_\lambda g_{\mu\nu}=\nabla_\lambda g_{\mu\nu}-\bbar{\Gamma}^\sigma_{\lambda\mu}g_{\sigma\nu}-\bbar{\Gamma}^\sigma_{\lambda\nu}g_{\mu\sigma}
\end{equation}
in the absence of torsion, such that $\nabla_\mu$ corresponds to the Levi-Civita connection and
\begin{equation}\label{WeylConnCoeff}
	\bbar{\Gamma}^\lambda_{\mu\nu}=-\frac{1}{2}g^{\lambda\sigma}(A_\mu g_{\nu\sigma}+A_\nu g_{\mu\sigma}-A_\sigma g_{\mu\nu})
\end{equation}
is the disformation for the Weyl connection.

The Weyl vector field can admit, in its most general form, a gauge group and a field strength that is given by
\begin{equation}\label{def:Weyl-StrengthTensor}
	F_{\mu\nu}=\partial_\mu A_\nu-\partial_\nu A_\mu-[A_\mu,A_\nu],
\end{equation}
where $[\cdot,\cdot]$ represents the commutator.
Naturally, if $A_\mu$ is an abelian field, the second term in the previous expressions vanishes.

The Lagrangian density of the vector field is given by:
\begin{equation}\label{def:YM-Lagrangian}
	\Lag_{\rm W}[A_\mu,g^{\mu\nu}] =-\frac{1}{4\mu}\tr{F_{\mu\nu}F^{\mu\nu}}-V[A],
\end{equation}
where $\mu$ is equivalent to the electromagnetic permeability, and a potential is admitted.
The associated energy-\linebreak-momentum tensor is obtained from the usual definition,
\begin{equation}
	T_{\mu\nu}^{\rm W}=-\frac{2}{\sqrtg}\frac{\delta(\sqrtg \Lag_{\rm W})}{\delta g^{\mu\nu}}.
\end{equation}
This leads to the Weyl energy-momentum tensor
\begin{align}
	T_{\mu\nu}^{\rm W} =\frac{1}{\mu}&\left(g^{\alpha\beta}\,\tr{F_{\mu\alpha}F_{\nu\beta}} -\frac{1}{4}g_{\mu\nu}\,\tr{F_{\alpha\beta} F^{\alpha\beta}}\right)\nonumber\\
	&\hspace{1em}-\left(g_{\mu\nu}-2\frac{\delta}{\delta g^{\mu\nu}}\right)V[A],
\end{align}
whose trace is given just by the second term
\begin{equation}\label{def:Trace-Tmunu}
	T^{\rm W}=g^{\mu\nu}T_{\mu\nu}^{\rm W} =-2\left(2-g^{\mu\nu}\frac{\delta}{\delta g^{\mu\nu}}\right)V[A].
\end{equation}
\pagebreak
From Eqs. \eqref{WeylConnection-LCConnection} and \eqref{WeylConnCoeff}, the Riemann tensor can be obtained.
Contracting this tensor, we get the Ricci tensor:
\begin{equation}
	\bar{R}_{\mu\nu}=R_{\mu\nu}+\frac{1}{2}g_{\mu\nu}\left(\nabla_\lambda-A_\lambda\right)A^\lambda-\frac{3}{2}A_\mu A_\nu-F_{\mu\nu}+\frac{1}{2}E_{\mu\nu},
\end{equation}
with $R_{\mu\nu}$ the Ricci tensor for the Levi-Civita connection; $E_{\mu\nu}=\nabla_\mu A_\nu+\nabla_\nu A_\mu+2\left\{A_\mu,A_\nu\right\}$ and $\{\cdot,\cdot\}$ representing the anti-commutator.
Finally, the scalar curvature is given by:
\begin{equation}
	\bar{R}=R+3\nabla^\lambda A_\lambda-\frac{3}{2}A^\lambda A_\lambda,
\end{equation}
where $R$ is the scalar curvature corresponding to the Levi-Civita connection.

\section{Non-minimally Coupled Weyl Connection Gravity}

The proposed NMCWCG theory has the following action:
\begin{align}
	S&=\int_M\left(\kappa f_1(\bar{R}) +f_2(\bar{R})\Lag_m\right)\sqrtg\,\ed^4x\\
	&-\int_Mf_2(\bar{R})\left(\frac{1}{4\mu}\tr{F_{\mu\nu}F^{\mu\nu}}+V[A]\right)\sqrtg\,\ed^4x,\nonumber
\end{align}
where $\kappa=\frac{c^4}{16\pi G}$ and $g$ is the determinant of the metric.
The equations of motion are given by:
\begin{widetext}
\begin{equation}\label{EOM:Metric}
	\left(R_{\mu\nu}+\frac{1}{2}A_{(\mu}A_{\nu)}+\frac{1}{2}g_{\mu\nu}(\nabla_\lambda-A_\lambda)A^\lambda+\nabla_{(\mu}A_{\nu)}\right)\bar{\Theta}-\frac{1}{2}f_1(\bar{R})g_{\mu\nu}=\frac{f_2(\bar{R})}{2\kappa}(T_{\mu\nu}^{(m)}+T_{\mu\nu}^{\rm W}),
\end{equation}
where $\bar{\Theta}=F_1(\bar{R})+\frac{1}{\kappa}\Lag\,F_2(\bar{R})$, $\Lag$ is the total Lagrangian (matter plus Weyl) and $F_i=\frac{\ed f_i}{\ed \bar{R}}$, $i=1,2$.
The corresponding trace equation is given by:
\begin{equation}\label{EOM:Trace-metric}
	\bar{R}\bar{\Theta}-2f_1(\bar{R})=\frac{f_2(\bar{R})}{2\kappa}(T^{(m)}+T^{\rm W}),
\end{equation}
where $T^{(m)}=g^{\mu\nu}\,T_{\mu\nu}^{(m)}$.
The equations of motion for the Weyl vector field are given by:
\begin{align}\label{EOM:Weyl-vector}
	\frac{1}{\mu}\nabla_\nu\big(f_2(\bar{R})F^{\mu\nu}\big)-\frac{f_2(\bar{R})}{\mu}[A_\nu,F^{\mu\nu}]=f_2(\bar{R})\frac{\delta(\Lag_m-V)}{\delta A_\mu}-3\kappa\big[\nabla^\mu\bar{\Theta}+A^\mu\bar{\Theta}\big],
\end{align}
where we have assumed that the Lagrangian of the matter fields depends on $A_\mu$ only through the covariant derivative $D_\mu$.
If we drop the kinetic term of the Weyl field and assume that the Lagrangian of the matter fields does not depend explicitly on $A_\mu$, we get the equations of motion obtained in Ref.~\cite{Gomes-Bertolami}.
\newpage
\end{widetext}

\section{Space-form}

A Riemannian manifold is said to be a space-form if its sectional curvature is constant, i.e. if the Riemann tensor components can be written as
\begin{equation}
	g_{\alpha\beta}\Riem[\bar{R}]{\alpha}{\gamma\delta\varepsilon}=\Lambda(g_{\beta\delta}g_{\gamma\varepsilon}-g_{\beta\gamma}g_{\delta\varepsilon}),
\end{equation}
with $\Lambda\in\mathbb{R}$.
A natural generalisation of this result, is to allow the space-form constant to vary with the coordinates of the manifold, $\Lambda=\Lambda(x)$.
Hence, we can compute the Ricci tensor and the scalar curvature as
\begin{subequations}\label{SF:Riemann}
\begin{align}
	\bar{R}_{\mu\nu}&=3\Lambda(x)g_{\mu\nu};\\[0.5em]
	\bar{R}&=12\Lambda(x).
\end{align}
\end{subequations}
It is easily seen, using Bianchi's Identity, that, for the Levi-Civita connection, the space-form must be a constant.
In the case of the Weyl connection, the previous identity gives us a constraint for the space-form function:
\begin{equation}
	\partial_\mu\Lambda(x)=A_\mu \Lambda(x).
\end{equation}

Substituting Eqs. \eqref{SF:Riemann} into Eq.~\eqref{EOM:Trace-metric} for\footnote{
	This form of the matter Lagrangian comes from considering the vacuum energy computed from quantum field theory.%
} $\Lag_m=-2\kappa\Lambda_0$ and $V[A]=0$ yields
\begin{align}
	&12\Lambda\left[F_1(12\Lambda)+\frac{F_2(12\Lambda)}{\kappa}\left(-2\kappa\Lambda_0-\frac{1}{4\mu}\tr{F_{\alpha\beta} F^{\alpha\beta}}\right)\right]\nonumber\\
	&\hspace{4em}=2f_1(12\Lambda)+\frac{f_2(12\Lambda)}{2\kappa}\big(-8\kappa\Lambda_0\big),
\end{align}
which, by simplifying and defining $\mathcal{F}\equiv\frac{1}{8\kappa\mu\Lambda_0}F_{\alpha\beta}F^{\alpha\beta}$, becomes
\begin{align}\label{EOM:SF:TracEq}
	6\Lambda&\left[F_1(12\Lambda)-2\Lambda_0\,F_2(12\Lambda)(1+\mathcal{F})\right]\nonumber\\
	&\hspace{4em}=f_1(12\Lambda)-2\Lambda_0f_2(12\Lambda),
\end{align}
where we have assumed that the Weyl field 
$A_\mu$
is constant.
This follows from setting the contribution of matter fields to be their non-vanishing vacuum expectation values.
Also notice that, because we are assuming a non-abelian Weyl field with field strength given by Eq.~\eqref{def:Weyl-StrengthTensor}, the quantity $\mathcal{F}$ does not vanish.
Furthermore, even though we have assumed that $V[A]=0$ to simplify the calculations, we could have also assumed $V[A]$ to be quadratic in the metric or quartic in the Weyl field.
In fact, potential-like terms arise from symmetry conditions as discussed in Refs.~\cite{Bento:1992wy,Bertolami:2015wir}.

Eq.~\eqref{EOM:SF:TracEq} can only be solved in terms of $\Lambda$ if we specify $f_1$ and $f_2$.
For that reason, we choose
\begin{equation}\label{f1-proposal}
	f_1(\bar{R})=\bar{R}+a\bar{R}^2,\ a\in\mathbb{R}\backslash\{0\}
\end{equation}
and consider the following cases \cite{Bertolami-2009,Bertolami-2010,Bertolami-2011.1,Bertolami-2011.2,Bertolami-2012,Bertolami-2015}:
\begin{subequations}\label{f2-proposals}
\begin{flalign}
	f_2(\bar{R})&=1+(\bar{R}/R_0)^n,\ n\in\mathbb{Z}\backslash\{0\};\label{f2-a}&&\\[0.5em]
	f_2(\bar{R})&=1+r_2 \bar{R}^2+r_{-1}/\bar{R},\ r_2,\,r_{-1}\in\mathbb{R}\backslash\{0\};&&\\[0.5em]
	f_2(\bar{R})&=1+\sum_{n\in U}r_n \bar{R}^n,\ U\subseteq\mathbb{Z}\backslash\{0\},\ r_n\in\mathbb{R}\backslash\{0\};\raisetag{2.45em}&&\\[0.5em]
	f_2(\bar{R})&=\exp(\bar{R}/R_0),\ R_0\in\mathbb{R}\backslash\{0\},&&
\end{flalign}
\end{subequations}
where 
$R_0,\,r_{-1},\,r_2$ and $r_n$ are reference curvatures
whose scale is given by the problem at hand, that is, it can be galactic, cluster or the observable horizon scale.

There are two major reasons for the choice of $f_1$: the first one is that this is the model proposed by Starobinsky for inflation \cite{Starobinsky}, which is consistent with the results from \textit{Planck} data 2018 \cite{Planck2018-Inflation};
the other reason is that, in Eq.~\eqref{EOM:SF:TracEq}, the only dependence on $f_1$ is $\bar{R}\,F_1(\bar{R})-2f_1(\bar{R})$ which vanishes for a quadratic term.
Thus, the introduction of a second order term in $f_1$ does not change the form of Eq.~\eqref{EOM:SF:TracEq} when compared to the equivalent expression with only the linear term of GR.
Of course, these considerations are valid as far as $f_2(\bar{R})$ does not affect the inflationary dynamics (see, however, Ref.~\cite{Gomes-Rosa-Bertolami-2018}).

Before substituting the proposed expressions for $f_1$ and $f_2$ (c.f. Eqs.~\eqref{f1-proposal} and \eqref{f2-proposals}) on Eq.~\eqref{EOM:SF:TracEq} and solve the resulting expression, we discuss the method to be employed.
The main focus in the next subsections will be to solve the trace equation, and check the possible values of the parameters arising from the functions $f_1$ and $f_2$, as well as the value of $\mathcal{F}$.
If solutions for values of $\Lambda$ as small as we want are found, we can then construct a cosmology built from this theory, since this procedure is equivalent to fine-tuning the cosmological constant in General Relativity.

The possible values for the parameters of the model are considered to be acceptable solutions of the system if they do not result in a divergent or trivial $f_2$ or a divergent value for the Weyl field vacuum expectation value.
Therefore, in the next subsections we search for values of $\Lambda\simeq0$ such that $R_0\centernot\to0$ or $R_0,\,r_{-1},\,r_2,\,\mathcal{F}\centernot\to\pm\infty$.
\newpage
\subsection{Power Coupling: $f_2(\bar{R})=1+\big(\bar{R}/R_0\big)^n$}
\label{SubSec:SF:PowerCoup}

We start by exploring the case of a power coupling,\linebreak Eq.~\eqref{f2-a}, where we are able to obtain an algebraic equation that relates $\Lambda$ and $\Lambda_0$ from Eq.~\eqref{EOM:SF:TracEq}.
The trace equation becomes
\begin{align}
	&6\Lambda\left[1+2a(12\Lambda)-2n\Lambda_0\frac{(12\Lambda)^{n-1}}{R_0^n}(1+\mathcal{F})\right]\\
	&\hspace{2em}-12\Lambda-a(12\Lambda)^2=-2\Lambda_0\left(1+\left(\frac{12\Lambda}{R_0}\right)^n\right)\nonumber.
\end{align}
Rearranging the terms, we get the equation
\begin{equation}\label{SF:Algebraic-Eq_Lambda-Lambda_0}
	\frac{12^n}{2R_0^n}\Lambda^n\big[n\mathcal{F}+(n-2)\big]+\frac{3}{\Lambda_0}\Lambda-1=0.
\end{equation}
This is a polynomial equation of type $p(\Lambda)=0$ for\linebreak $p(\Lambda)=\alpha\Lambda^n+\beta\Lambda-1$. It is easily shown, using Rolle's theorem, that this type of polynomial has at most two or three real roots depending on the sign and parity of $n$.

We can now solve Eq.~\eqref{SF:Algebraic-Eq_Lambda-Lambda_0} for some values of $n$:
\begin{subequations}\label{PowerCoup-itemize}
\begin{itemize}
	\item[] $n=2$:
	\begin{equation}
		\Lambda=\frac{R_0^2}{\Lambda_0}\left(\frac{-1\pm\sqrt{1+64\Lambda_0^2\mathcal{F}/R_0^2}}{96\mathcal{F}}\right),
	\end{equation}
	\item[] $n=1$:
	\begin{equation}
		\Lambda=\frac{\Lambda_0/3}{1+2\Lambda_0(\mathcal{F}-1)/R_0}.
	\end{equation}
	\item[] In the case of $n=-1$ we have to invert the equation. Therefore, Eq.~\eqref{SF:Algebraic-Eq_Lambda-Lambda_0} becomes
	\begin{equation}
		-\frac{R_0}{24}(\mathcal{F}+3)+\frac{3\Lambda^2}{\Lambda_0}-\Lambda=0,
	\end{equation}
	whose solutions are:
	\begin{equation}
		\Lambda=\frac{\Lambda_0}{6}\left(1\pm\sqrt{1+\frac{R_0}{2\Lambda_0}(\mathcal{F}+3)}\right).
	\end{equation}
\end{itemize}
\end{subequations}
Notice that for all cases, in order to have a solution in which $\Lambda/\Lambda_0\to0$ we must have $R_0/\Lambda_0\to 0$ (for $n=2$, $\vert R_0\vert\to\infty$ also yields the same effect).
For the positive values of $n$ we can also obtain $\Lambda/\Lambda_0\to0$ by allowing $\mathcal{F}\to\infty$.
Clearly, this last choice is not physically acceptable.
For $n=-1$, even though the choice $\mathcal{F}=-3$ seems to give a flat space-time solution, we have multiplied the whole equation by $\Lambda$, which invalidates this solution.
Thus, for this case, we can only get $\Lambda/\Lambda_0\simeq0$, when $\mathcal{F}\simeq-3$.
When we divide the solutions of Eqs.~\eqref{PowerCoup-itemize} by $R_0$ and check for $R_0/\Lambda_0\to0$, we find that, for $n=2$ and $n=-1$, $\Lambda/R_0\to0$ and $\Lambda/R_0\to\infty$, respectively.
These are, again, not physically acceptable since they correspond to $f_2=1$ and $f_2\to\infty$.
For $n=1$, we get the constraint $6\Lambda(\mathcal{F}-1)\simeq R_0$.
Hence, if $\Lambda$ is small, $R_0\to0$ or $\mathcal{F}\to\infty$ which are not acceptable.

For values of $n$ larger than 2 and smaller that $-1$, an analytical solution becomes difficult to get (and in most cases impossible due to the Abel-Ruffini theorem).
Therefore, we shall try the approximation\footnote{This approximation is valid in the sense that, observationally, $\Lambda\lesssim 10^{-52}\,\mathrm{m}^{-2}$ \cite{Planck2018-Cosmology}, whilst the value of the vacuum energy given by quantum field theory is $\Lambda_0\sim10^{63}\,\mathrm{m}^{-2}$, when considering the vacuum energy density to be $\rho_{\rm vac}=\frac{\Lambda_0}{8\pi G}\sim 10^{71}\,\mathrm{GeV}^4$. Consequently, when comparing these values, $\Lambda/\Lambda_0\lesssim 10^{-115}$.}
$\Lambda\ll\Lambda_0$ directly in  Eq.~\eqref{SF:Algebraic-Eq_Lambda-Lambda_0}.
In order to do that, we drop the linear term and thus:
\begin{equation}
	\frac{12^n}{2R_0^n}\Lambda^n\big(n\mathcal{F}+(n-2)\big)\simeq 1.
\end{equation}
Now, the solution to this expression is trivial.
For $n$ odd:
\vspace{-0.5cm}
\begin{equation}
	\Lambda\simeq\frac{R_0}{12}\left(-1+\frac{n}{2}\big(\mathcal{F}+1\big)\right)^{-1/n};
\end{equation}
and for the case of $n$ even:
\begin{equation}
	\Lambda\simeq\pm\frac{R_0}{12}\left(-1+\frac{n}{2}\big(\mathcal{F}+1\big)\right)^{-1/n}.
\end{equation}
Again we see that for any $n$, $\Lambda$ is arbitrarily small for $R_0$ arbitrarily small or for $\mathcal{F}$ large for positive $n$ or 
\begin{equation}
	\mathcal{F}\simeq-1+\frac{2}{n}
\end{equation}
for $n$ negative.
Thus, the power coupling only yields acceptable solutions for negative values of $n$.

\subsection{Power Series: $f_2(\bar{R})=1+\sum_{n\in U}r_n \bar{R}^n$}

The next test is to consider $f_2$ to be a finite sum of several powers.
For that reason, let $U$ be a subset of $\mathbb{Z}\backslash\{0\}$ so that
\begin{equation}
	f_2(\bar{R})=1+\sum_{n\in U} r_n\, \bar{R}^n,\quad r_n\neq 0\hspace{0.75em}\forall\, n\in U.
\end{equation}
Then, Eq.~\eqref{EOM:SF:TracEq} becomes
\begin{align}
	3\Lambda&\left(1+2\Lambda_0(1+\mathcal{F})\sum_{n\in U}nr_n (12\Lambda)^{n-1}\right)\nonumber\\
	&\hspace{3em}=\Lambda_0\left(1+\sum_{n\in U}r_n (12\Lambda)^n\right),
\end{align}
which leads to,
\begin{equation}\label{SF:Eq:arbit-U}
	\sum_{n\in U}r_n 12^n\Lambda^n\left(\frac{n}{2}(1+\mathcal{F})-1\right)=1-3\frac{\Lambda}{\Lambda_0}.
\end{equation}

Obviously, we cannot solve this equation without knowing the full set $U$. Let us then consider $U=\{-1,2\}$.
In this case, Eq.~\eqref{SF:Eq:arbit-U} becomes
\begin{equation}\label{SF:Eq:U={-1,2}}
	\frac{r_{-1}}{ 12\Lambda}\left(-\frac{1}{2}(1+\mathcal{F})-1\right)+144r_2\Lambda^2\mathcal{F}=1-3\frac{\Lambda}{\Lambda_0}.
\end{equation}
Multiplying the whole equation by $\Lambda$, we are able to find the expression
\begin{equation}
	(144 r_2\mathcal{F})\Lambda^3 +\left(\frac{3}{\Lambda_0}\right)\Lambda^2-\Lambda-\frac{r_{-1}}{24}\big(\mathcal{F}+3\big)=0.
\end{equation}

Because, computationally, it is easier to work with dimensionless quantities, we divide the equation by $\Lambda_0$, leading to:
\begin{equation}\label{SF:-1,2_Algebric-Equation}
	\left(144\rho_2\mathcal{F}\right)\lambda^3+3\lambda^2-\lambda-\frac{\rho_{-1}}{24}\big(\mathcal{F}+3\big)=0,
\end{equation}
where $\rho_2\equiv r_2\Lambda_0^2$, $\rho_{-1}\equiv r_{-1}/\Lambda_0$ and $\lambda\equiv\Lambda/\Lambda_0$.
From a simple analysis we see that, if we set $\rho_2=0$ or $\mathcal{F}=0$, we reobtain the solution for $n=-1$ from the previous section.
If, instead, we assume\footnote{One could say that we could also set $\rho_{-1}=0$. That is not the case, since to get to Eq.~\eqref{SF:-1,2_Algebric-Equation}, we had to assume $r_{-1}\neq 0$.}
that $\mathcal{F}=-3$, we get the solutions of $n=2$ of the previous section, as well as the solution $\lambda=0$.
Although this is exactly the result we would want, $\Lambda=0$ is not an acceptable solution when considering Eq.~\eqref{SF:Eq:U={-1,2}}.
Notice that the choice of a negative power and a positive power are motivated by the fact that a negative power tends to mimic dark energy \cite{Bertolami-2010}, whilst a positive power is useful to mimic dark matter \cite{Bertolami-2012,Bertolami-2011.1,Bertolami-Paramos}.

\pagebreak
We can also solve Eq.~\eqref{SF:-1,2_Algebric-Equation} numerically.
The graphs in Figs.~\ref{fig:Series-Coup-rho-1} and \ref{fig:Series-Coup-rho2} represent the solutions depending on the parameters $\rho_{-1}$, $\rho_2$ for $\mathcal{F}=7$.
In Figs.~\ref{fig:Series-Coup-F_fixedrho-1} and \ref{fig:Series-Coup-F_fixedrho2}, we plot the solutions of Eq.~\eqref{SF:-1,2_Algebric-Equation} for any $\mathcal{F}$ for different values of $\rho_2$ with $\rho_{-1}=2$ and for different values of $\rho_{-1}$ and $\rho_2=2$, respectively.

We find that, for a constant value $\mathcal{F}$, the only possible way to get $\lambda\simeq 0$ is when either $\vert\rho_{-1}\vert$ or $\vert\rho_2\vert$ are large, which are not interesting results (they both correspond to the case $f_2\to\pm\infty$).
However, for varying $\mathcal{F}$ fine-\linebreak-tuning is achievable.
By analysing Figs.~\ref{fig:Series-Coup-F_fixedrho-1} and \ref{fig:Series-Coup-F_fixedrho2}, we find solutions where $\Lambda\simeq 0$ for values $\mathcal{F}\simeq-3$.
This means that, for
\begin{equation}
	\frac{1}{4\mu}F_{\alpha\beta}F^{\alpha\beta}\simeq -6\kappa\Lambda_0
\end{equation}
or, for
\begin{equation}
	\langle\Lag_{\rm W}\rangle_0\simeq -3\langle\Lag_m\rangle_0,
\end{equation}
where $\langle\,\cdot\,\rangle_0$ represents the vacuum expectation values, we find solutions for $\Lambda\simeq 0$.
\begin{widetext}
\begin{minipage}{0.47\textwidth}
\begin{figure}[H]
	\centering
	\includegraphics[width=0.95\textwidth]{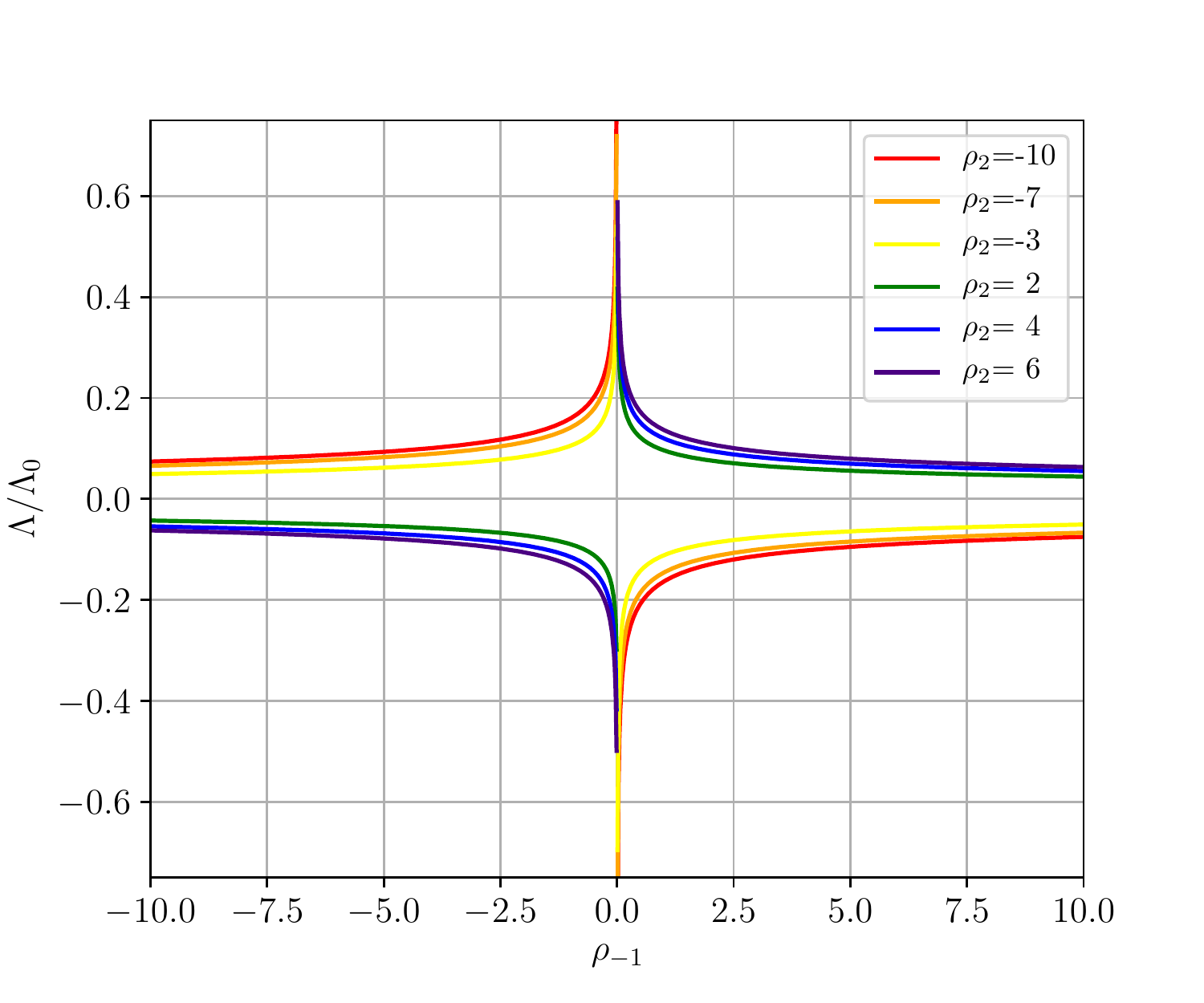}
	\caption{Solutions to Eq.~\eqref{SF:-1,2_Algebric-Equation} dependent on parameters $\rho_{-1}$ for $\mathcal{F}=7$ and for different values of $\rho_2$.}
	\label{fig:Series-Coup-rho-1}
\end{figure}
\end{minipage}\hspace*{.5cm}
\begin{minipage}{0.47\textwidth}
\begin{figure}[H]
	\centering
	\includegraphics[width=0.95\textwidth]{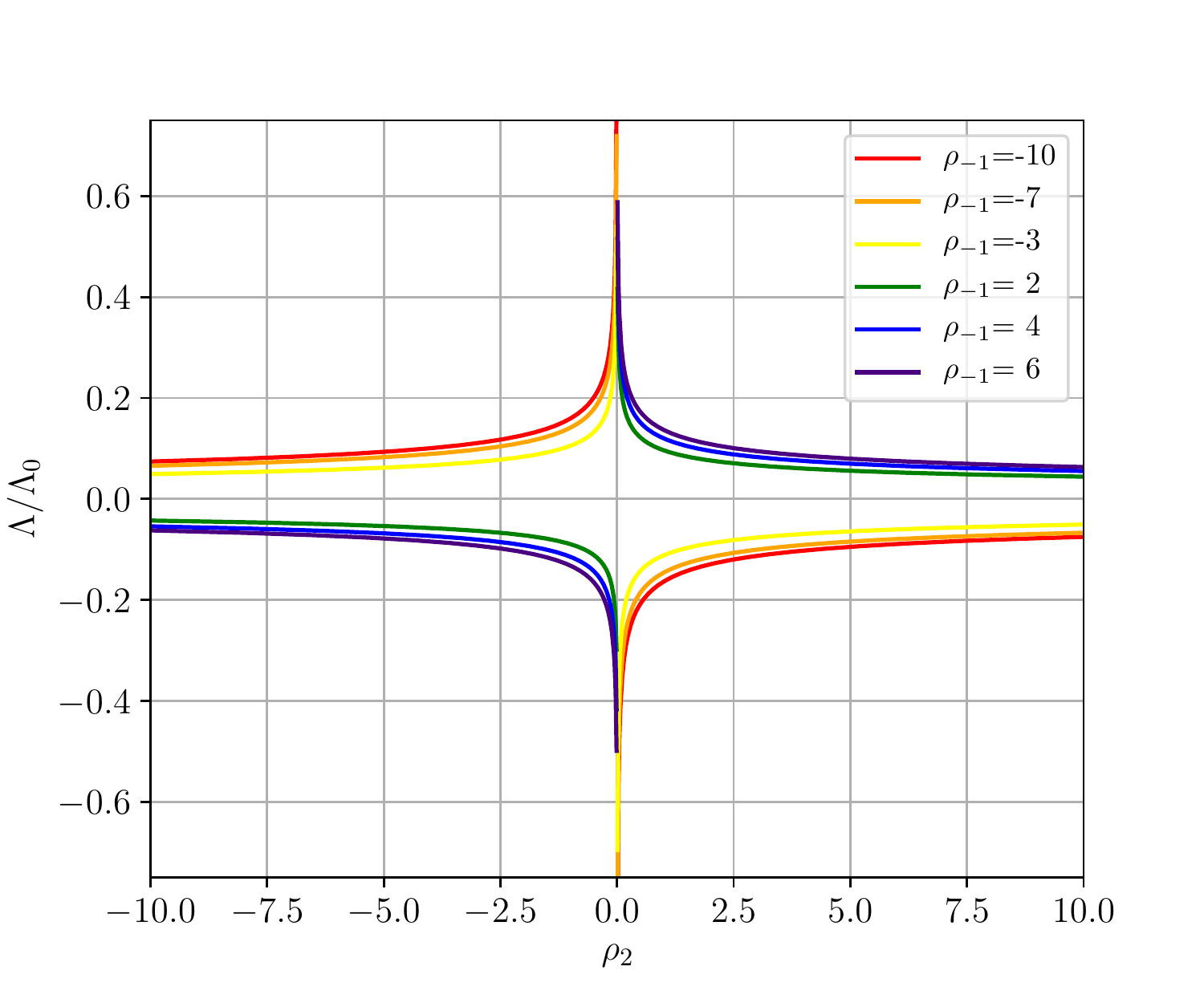}
	\caption{Solutions to Eq.~\eqref{SF:-1,2_Algebric-Equation} dependent on parameters $\rho_2$ for $\mathcal{F}=7$ and for different values of $\rho_{-1}$.}
	\label{fig:Series-Coup-rho2}
\end{figure}
\end{minipage}

\begin{minipage}{.47\textwidth}
\begin{figure}[H]
	\centering
	\includegraphics[width=0.95\textwidth]{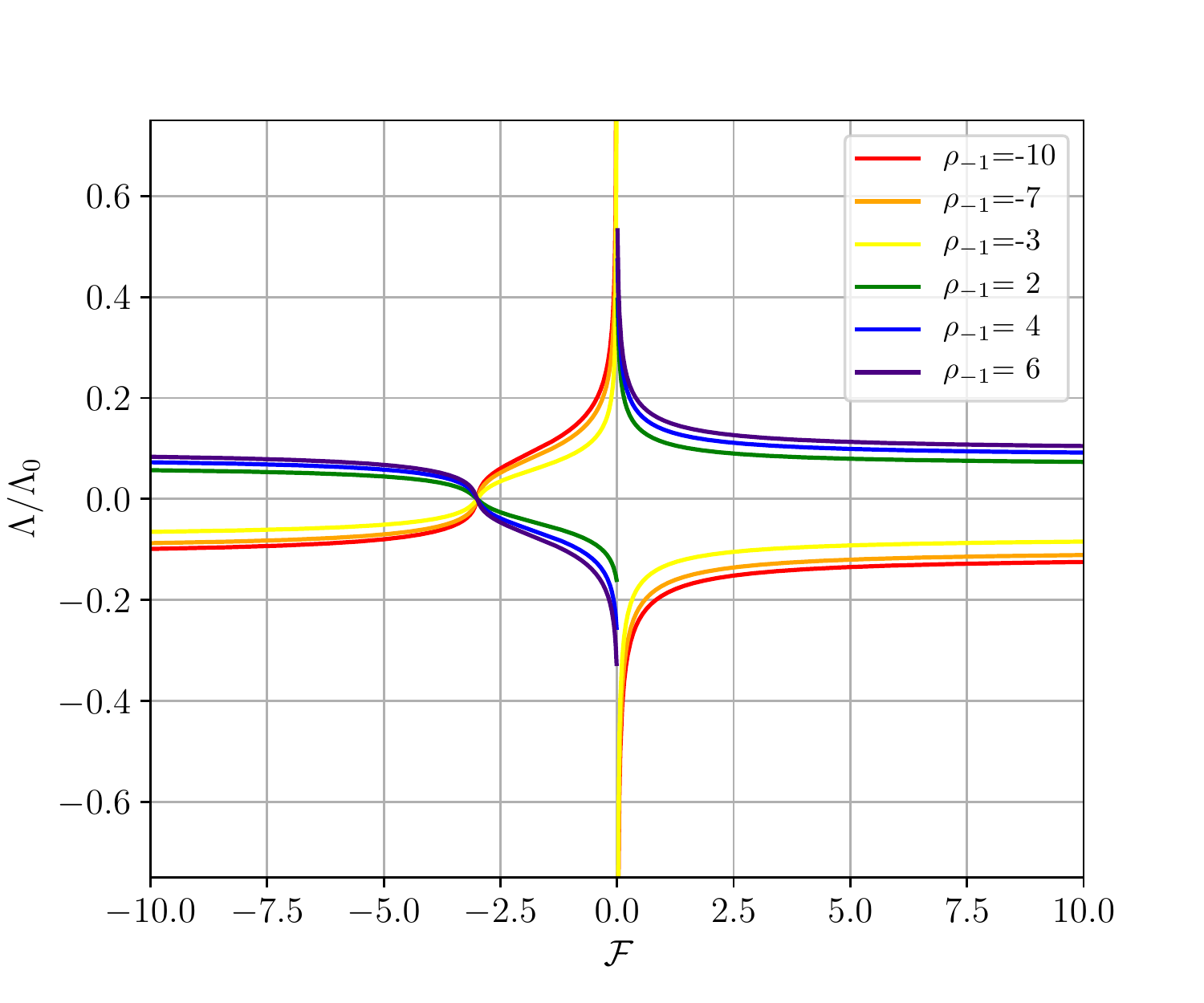}
	\caption{Solutions to Eq.~\eqref{SF:-1,2_Algebric-Equation} dependent on $\mathcal{F}$ for different values of $\rho_{-1}$ and for $\rho_2=2$.}
	\label{fig:Series-Coup-F_fixedrho-1}
\end{figure}
\end{minipage}\hspace*{.5cm}
\begin{minipage}{.47\textwidth}
\begin{figure}[H]
	\centering
	\includegraphics[width=0.95\textwidth]{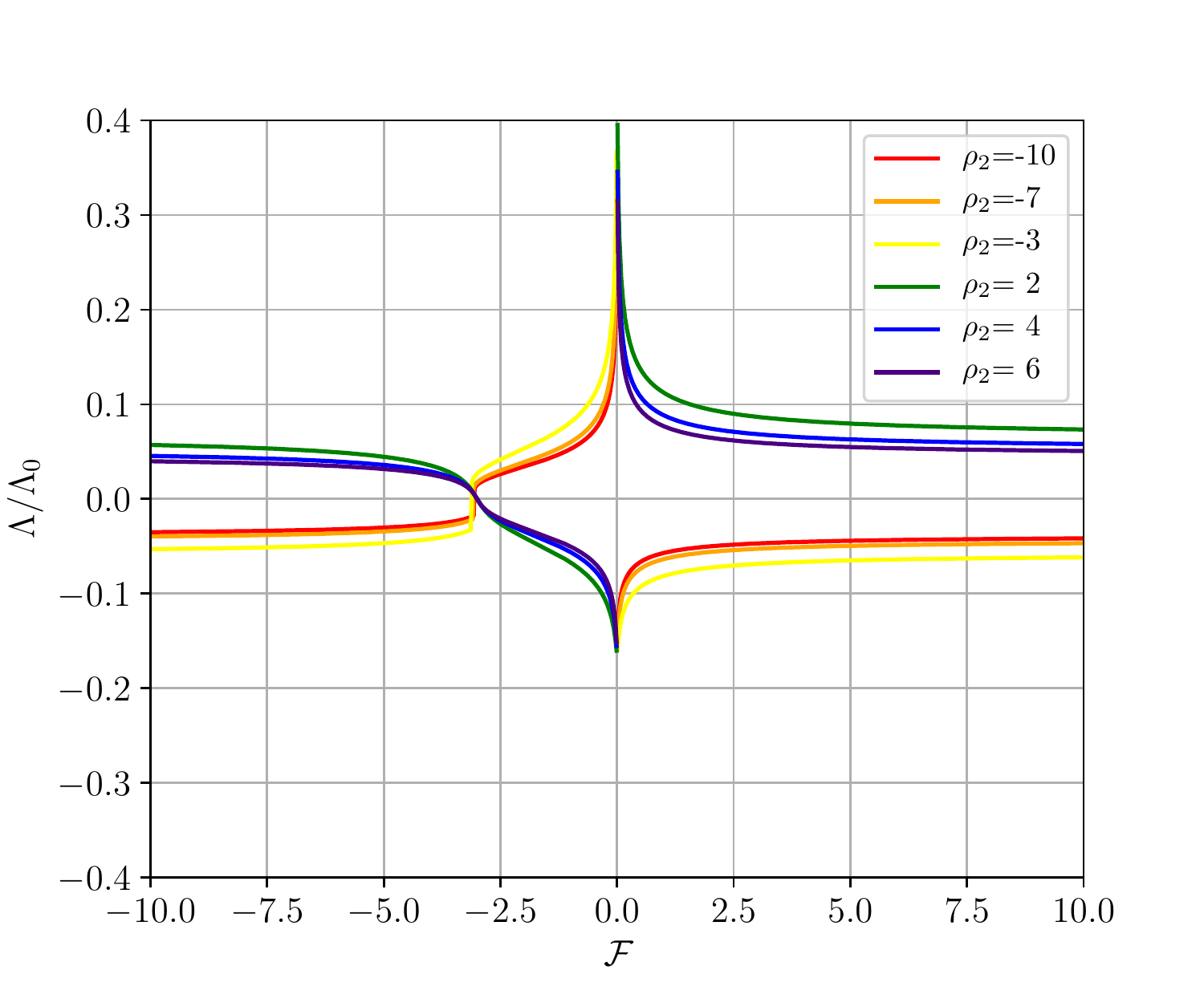}
	\caption{Solutions to Eq.~\eqref{SF:-1,2_Algebric-Equation} dependent on $\mathcal{F}$ for different values of $\rho_2$ and for $\rho_{-1}=2$.}
	\label{fig:Series-Coup-F_fixedrho2}
\end{figure}
\end{minipage}
\end{widetext}

\subsection{Exponential Coupling: $f_2(\bar{R})=e^{\bar{R}/R_0}$}

For this last case, we restart with the trace equation, Eq.~\eqref{EOM:Trace-metric}, which becomes
\begin{align}
	\bar{R}&\left(1+2a\bar{R}-\frac{1}{\kappa R_0}e^{\bar{R}/R_0}\left(2\kappa\Lambda_0+\frac{1}{4\mu}F_{\alpha\beta}F^{\alpha\beta}\right)\right)\nonumber\\[1em]
	&\hspace{2em}=2(\bar{R}+a\bar{R}^2)-4\Lambda_0 e^{\bar{R}/R_0}.
\end{align}
Defining $r\equiv \bar{R}/\Lambda_0=12\Lambda/\Lambda_0$, $r_0\equiv R_0/\Lambda_0$ and, again, $\mathcal{F}\equiv\frac{1}{8\kappa\mu\Lambda_0}F_{\alpha\beta}F^{\alpha\beta}$, we obtain the transcendental equation
\begin{equation}\label{SF:Transc-EQ.}
	\frac{1}{4}\,r\,e^{-r/r_0}=1-\frac{1}{2}\frac{r}{r_0}\big(1+\mathcal{F}\big).
\end{equation}
It is easy to see from a graphical analysis of both sides of Eq.~\eqref{SF:Transc-EQ.} that there are, at most, three solutions.
Nevertheless, we are interested in the case $r\ll 1$, corresponding to $\Lambda\ll\Lambda_0$.
Hence, in Figs.~\ref{fig:ExpCoup-Fsmall-lambda_r}--\ref{fig:ExpCoup-Flarge-lambda:r_r}, we plot the numerical solutions of Eq.~\eqref{SF:Transc-EQ.} as a function of the free parameters $r_0$ and $\mathcal{F}$.
The graphs are organised as follows: the ones on the left (thus, Figs.~\ref{fig:ExpCoup-Fsmall-lambda_r} and \ref{fig:ExpCoup-Fsmall-lambda:r_r}) we plot the numerical solutions for $\mathcal{F}\leq-1$ whilst the graphs on the right (Figs.~\ref{fig:ExpCoup-Flarge-lambda_r} and \ref{fig:ExpCoup-Flarge-lambda:r_r}) show the ones for $\mathcal{F}>-1$.
Also, the graphs on top show the $r$ vs.$\,r_0$ plane whilst the graphs on the bottom show the $r/r_0$ vs. $r_0$ plane.

The plots of Figs.~\ref{fig:ExpCoup-Fsmall-lambda_r} and \ref{fig:ExpCoup-Flarge-lambda_r} show that the solutions of Eq.~\eqref{SF:Transc-EQ.} that satisfy $\Lambda\ll\Lambda_0$ appear only if $R_0/\Lambda_0\sim0$ for any value of $\mathcal{F}$.
Since the value of $\Lambda_0$ is considerably large, the value of $R_0$ does not necessarily need to vanish.
Thus, we may find solutions for $R_0$ by dividing the y-axis by the x-axis.
This results in graphs of the form $\Lambda/R_0$ as seen Figs.~\ref{fig:ExpCoup-Fsmall-lambda:r_r} and \ref{fig:ExpCoup-Flarge-lambda:r_r}.
These graphs show an inversely proportional dependence between $\Lambda/R_0$ and $\mathcal{F}$ for $R_0/\Lambda_0\sim0$.
To better understand this dependence we refer back to Eq.~\eqref{SF:Transc-EQ.}.
Redefining $\lambda\equiv\Lambda/R_0$, we may eliminate $r$ of Eq.~\eqref{SF:Transc-EQ.}:
\begin{equation}
	\frac{1}{4}12\lambda\,r_0\,e^{-12\lambda}=1-\frac{1}{2}12\lambda(1+\mathcal{F}),
\end{equation}
which, by isolating $r_0$ becomes
\begin{equation}\label{SF:Transc-ParSpace}
	r_0=e^{12\lambda}\left(\frac{1}{3\lambda}-2(1+\mathcal{F})\right).
\end{equation}

Since we are looking for values of $r_0\sim0$, we may write
\begin{equation}
	e^{12\lambda}\left(\frac{1}{3\lambda}-2(1+\mathcal{F})\right)\simeq0
\end{equation}
which correspond to the solutions
\begin{equation}
	\lambda\to-\infty\quad{\rm or}\quad\lambda\simeq\frac{1}{6(1+\mathcal{F})}
\end{equation}
or, by rewriting in terms of the initial quantities (and noticing that $\Lambda\centernot\to-\infty$),
\begin{equation}
	R_0\to 0^{-}\quad{\rm or}\quad\Lambda\simeq\frac{1}{6}\frac{R_0}{1+\mathcal{F}}.
\end{equation}
Since the solution on the left corresponds to a vanishing $f_2$, this solution is not acceptable.
The same is not true for the solution on the right which gives a constraint to the values of parameter space.

\begin{widetext}
\vspace{-1em}
\begin{minipage}{0.47\textwidth}
	\begin{figure}[H]
		\centering
		\includegraphics[width=0.95\textwidth]{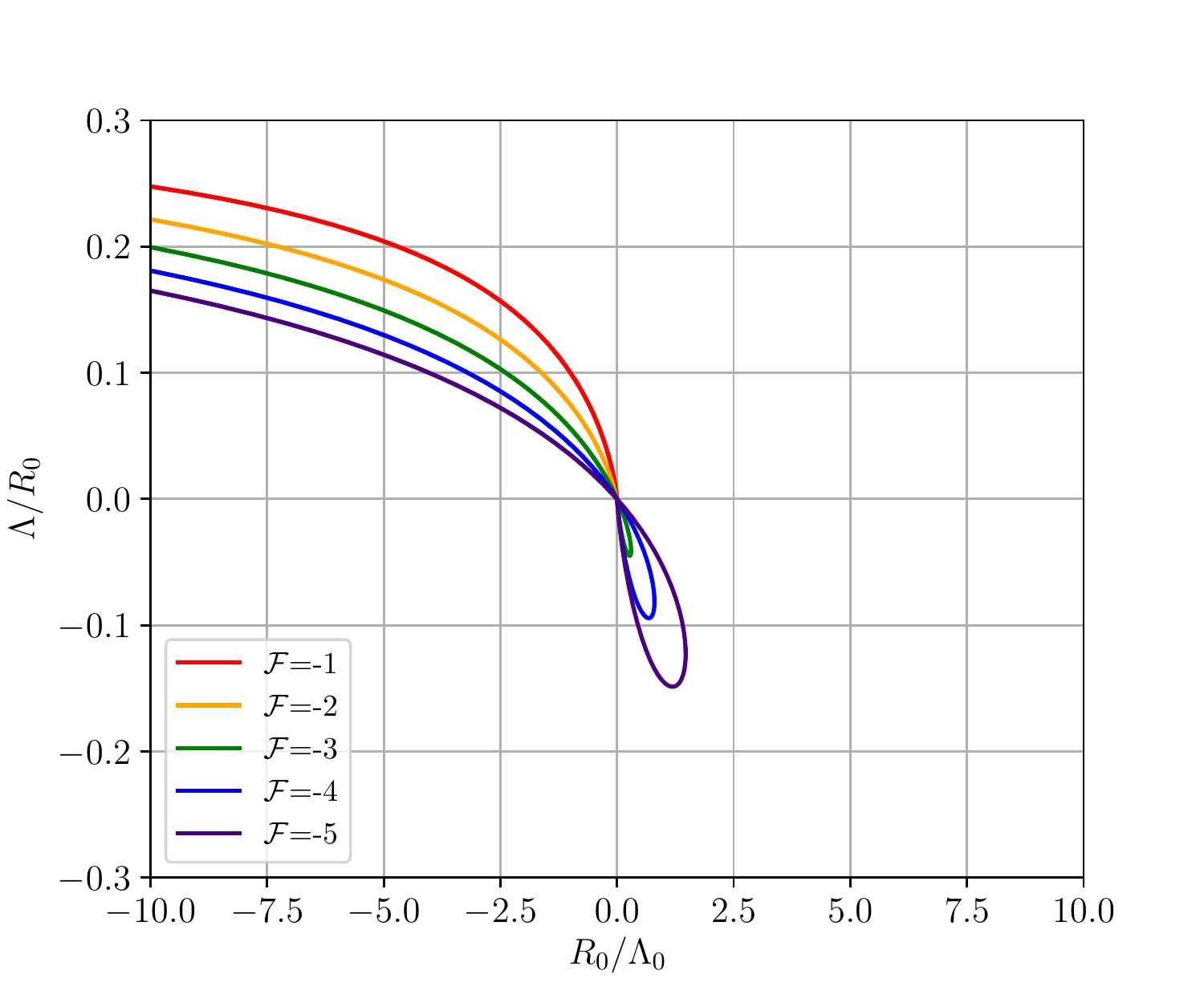}
		\caption{Solutions $\Lambda/\Lambda_0$ of Eq.~\eqref{SF:Transc-EQ.} for $\mathcal{F}\leq-1$.}
		\label{fig:ExpCoup-Fsmall-lambda_r}
	\end{figure}
\end{minipage}\hspace*{.5cm}
\begin{minipage}{0.47\textwidth}
	\begin{figure}[H]
		\centering
		\includegraphics[width=0.95\textwidth]{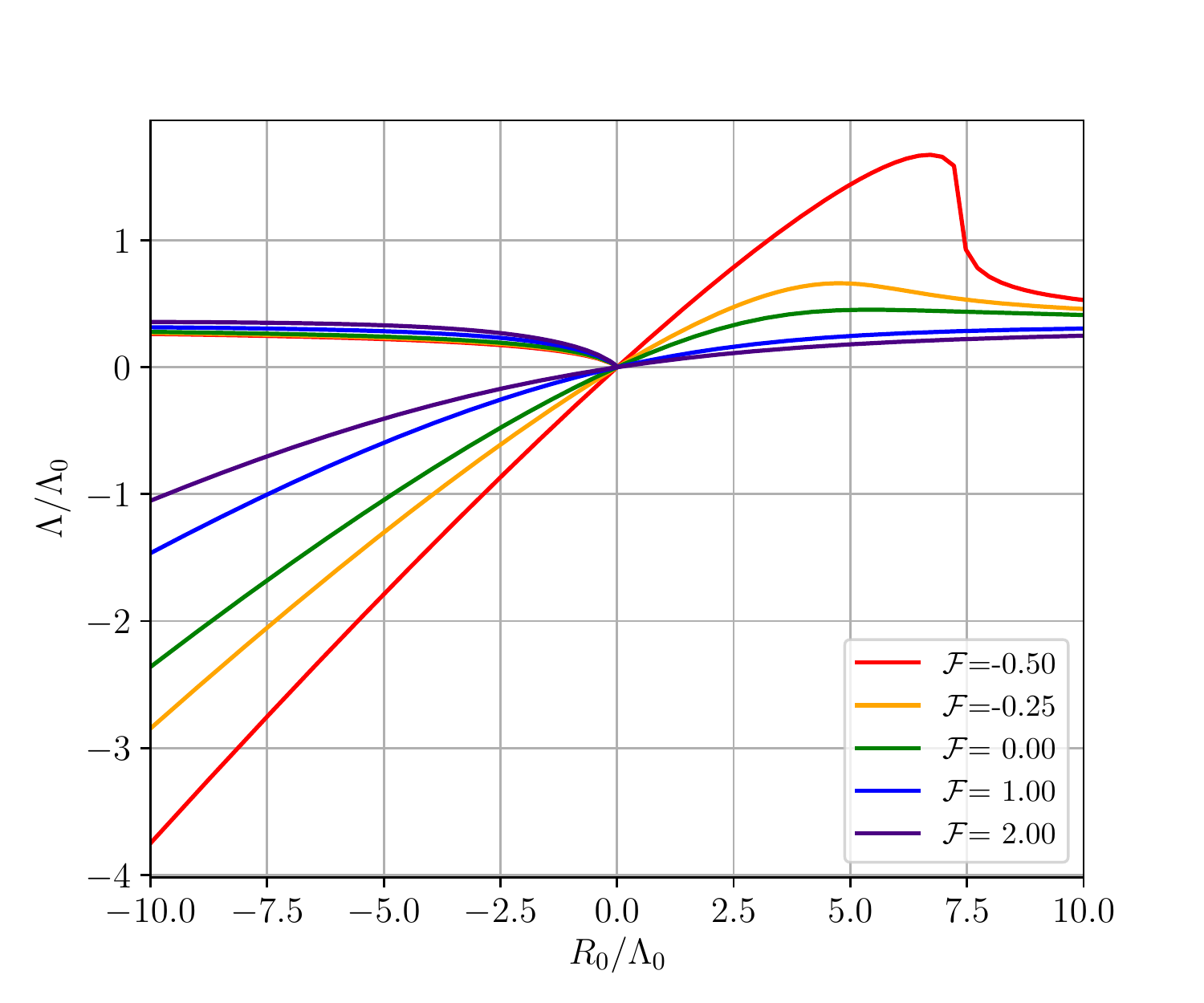}
		\caption{Solutions $\Lambda/\Lambda_0$ of Eq.~\eqref{SF:Transc-EQ.} for $\mathcal{F}>-1$.}
		\label{fig:ExpCoup-Flarge-lambda_r}
	\end{figure}
\end{minipage}

\begin{minipage}{.47\textwidth}
	\begin{figure}[H]
		\centering
		\includegraphics[width=0.95\textwidth]{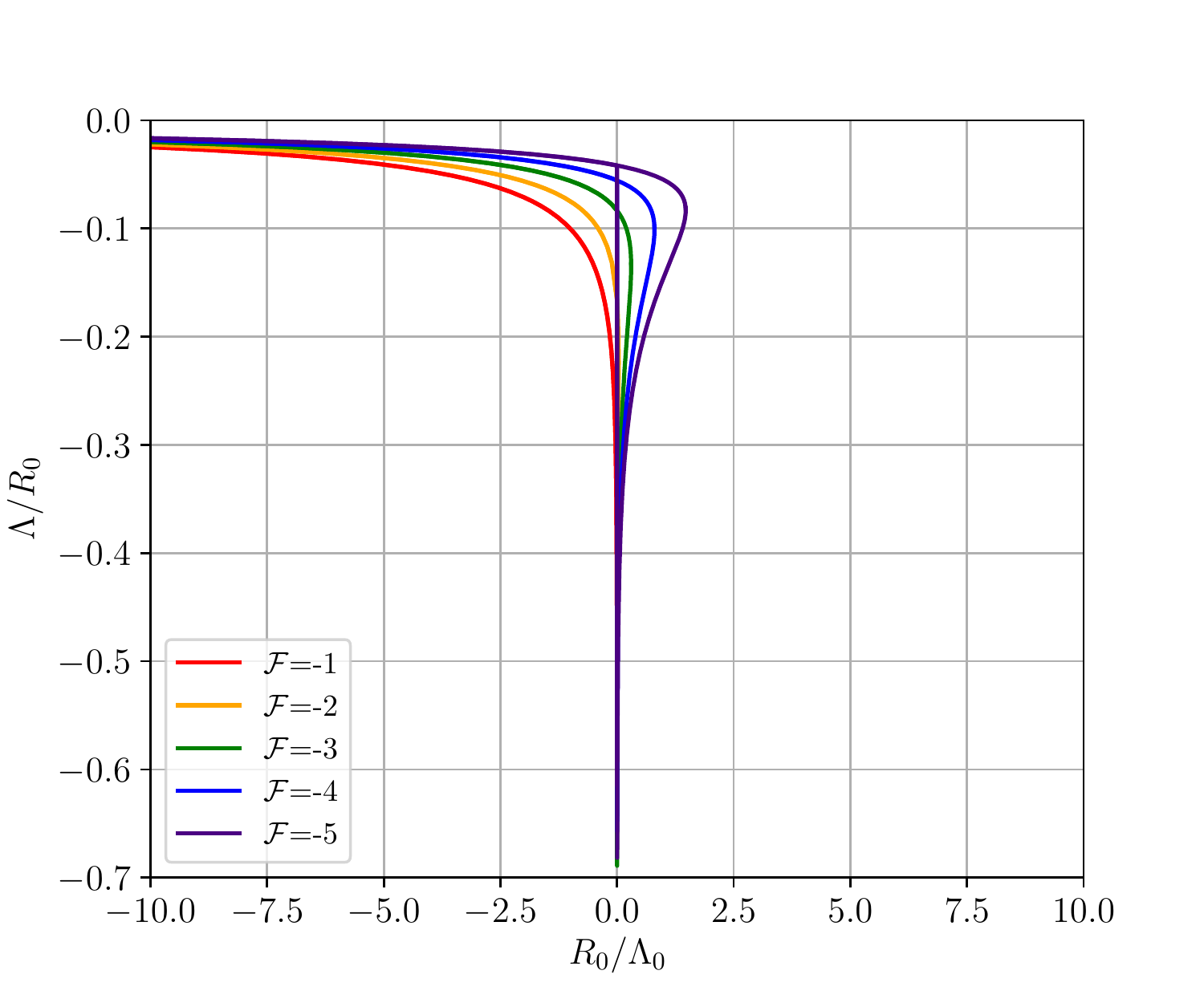}
		\caption{Solutions $\Lambda/R_0$ of Eq.~\eqref{SF:Transc-EQ.} for $\mathcal{F}\leq-1$.}
		\label{fig:ExpCoup-Fsmall-lambda:r_r}
	\end{figure}
\end{minipage}\hspace*{.5cm}
\begin{minipage}{.47\textwidth}
	\begin{figure}[H]
		\centering
		\includegraphics[width=0.95\textwidth]{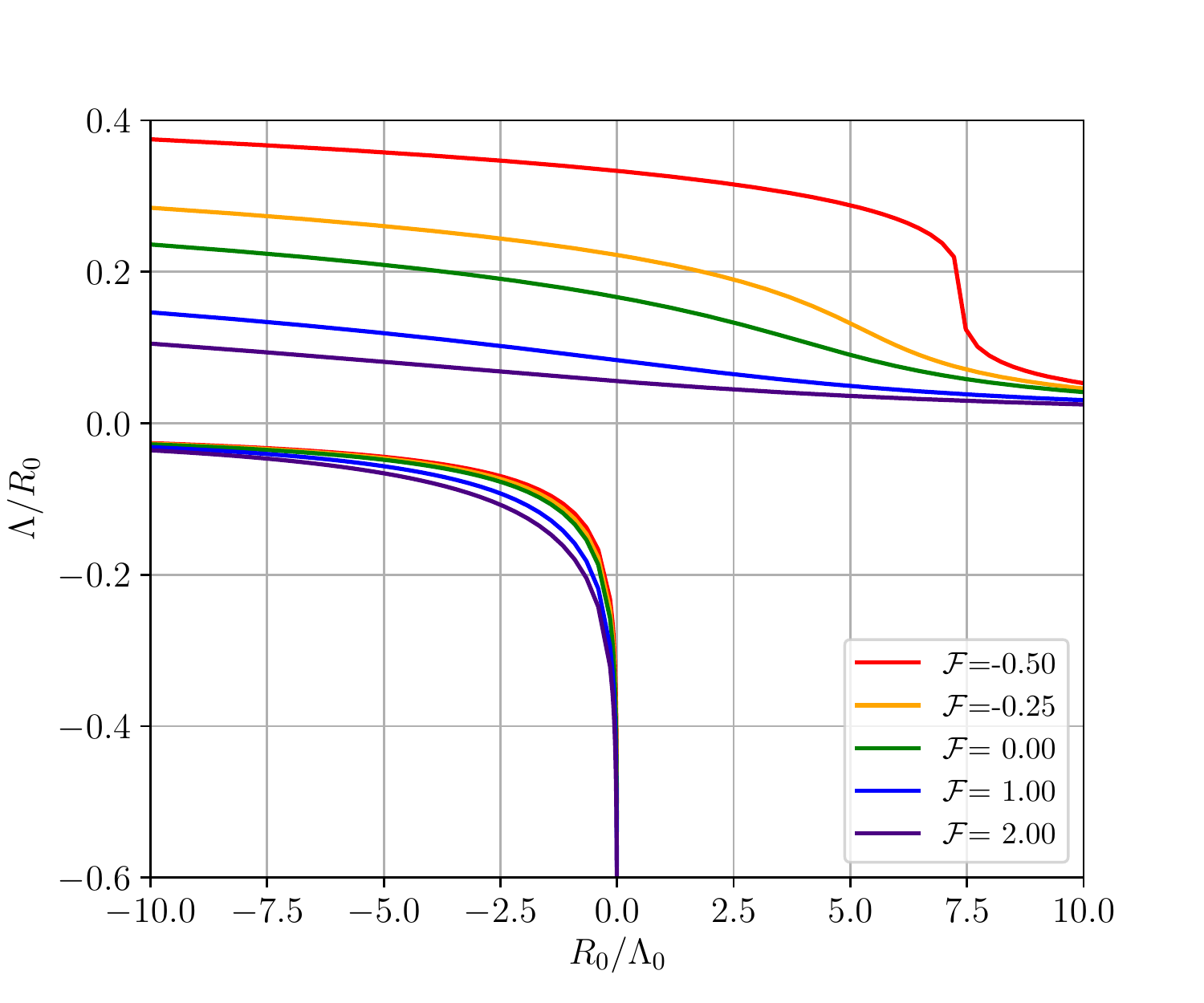}
		\caption{Solutions $\Lambda/R_0$ of Eq.~\eqref{SF:Transc-EQ.} for $\mathcal{F}>-1$.}
		\label{fig:ExpCoup-Flarge-lambda:r_r}
	\end{figure}
\end{minipage}
\end{widetext}

\section{Conclusions}

In this work we have considered the non-minimally coupled Weyl connection gravity (NMCWCG) where the Weyl vector field is dynamical.
We have examined solutions of the system for a space-form manifold.

We have looked for solutions with the choice $f_1(\bar{R})=\bar{R}+a\bar{R}^2$, and study different possibilities for the coupling, $f_2(\bar{R})$: an arbitrary power of $\bar{R}$, multiple powers of $\bar{R}$ and an exponential form.
The obtained results for values of a constant space-form small compared with the vacuum energy of matter fields are as follows:
for the power coupling with power $n$, we only find acceptable solutions for $n<0$, $\mathcal{F}=-1+2/n$, since for $n>0$ the corresponding parameters lead to a divergent term for the function $f_2$ or a trivial one.
For $f_2(\bar{R})=r_2\bar{R}^2+r_{-1}/\bar{R}$, if $\mathcal{F}\simeq-3$ or, similarly, for $\langle\Lag_{\rm W}\rangle_0\simeq-3\langle\Lag_m\rangle_0$.
Finally, for the exponential coupling, we find that, regardless of $\mathcal{F}$, the value of $R_0/\Lambda_0$ has to be small.
This final result prompted us to compare the values of $\Lambda$ and $R_0$ which leads to a constraint in the space of parameters $R_0,\,\mathcal{F}$, given by $\Lambda\simeq\frac{R_0/6}{1+\mathcal{F}}$.

These results show that the NMCWCG model can be considered for setting up cosmological models for certain choices of $f_2(\bar{R})$ as described above.
\newpage

\end{document}